\documentclass[amssymb,prb,aps,twocolumn,amsmath,showpacs,superscriptaddress]{revtex4}

\usepackage[dvips]{graphicx}

\renewcommand{\Im}{\mathop{\mathrm{Im}}}

\newcommand{\fixme}[1]{}

\begin{document}

\title{Nonequilibrium transport in mesoscopic multi-terminal SNS Josephson junctions}
\author{M.S. Crosser}

\affiliation{Department of Physics and Astronomy, Michigan State
University, East Lansing, MI 48824-2320, USA}
\affiliation{Department of Physics, Linfield College, 900 SE Baker
Street, McMinnville, OR 97128, USA}

\author{Jian Huang}

\altaffiliation[Present address: ]{Department of Physics, Taylor
University, Upland, IN  46989, USA} \affiliation{Department of
Physics and Astronomy, Michigan State University, East Lansing, MI
48824-2320, USA}

\author{F. Pierre}

\altaffiliation[Present address: ]{Laboratoire de Photonique et de
Nanostructures-CNRS, Route de Nozay, 91460 Marcoussis,
France}\affiliation{Department of Physics and Astronomy, Michigan
State University, East Lansing, MI 48824-2320, USA}

\author{Pauli Virtanen}

\author{Tero T. Heikkil\"{a}}

\affiliation{Low Temperature Laboratory, Helsinki University of
Technology, P. O. Box 2200, FIN-02015 TKK, Finland}

\author{F. K. Wilhelm}

\affiliation{Department of Physics and Astronomy and Institute for
Quantum Computing, University of Waterloo, Waterloo, Ontario, N2L
3G1, Canada}

\author{Norman O. Birge}

\email{birge@pa.msu.edu}

\affiliation{Department of Physics and Astronomy, Michigan State
University, East Lansing, MI 48824-2320, USA}

\date{\today}

\begin{abstract}
We report the results of several nonequilibrium experiments
performed on superconducting/normal/superconducting (S/N/S)
Josephson junctions containing either one or two extra terminals
that connect to normal reservoirs. Currents injected into the
junctions from the normal reservoirs induce changes in the electron
energy distribution function, which can change the properties of the
junction.  A simple experiment performed on a 3-terminal sample
demonstrates that quasiparticle current and supercurrent can coexist
in the normal region of the S/N/S junction. When larger voltages are
applied to the normal reservoir, the sign of the current-phase
relation of the junction can be reversed, creating a
``$\pi$-junction."  We compare quantitatively the maximum critical
currents obtained in 4-terminal $\pi$-junctions when the voltages on
the normal reservoirs have the same or opposite sign with respect to
the superconductors. We discuss the challenges involved in creating
a ``Zeeman" $\pi$-junction with a parallel applied magnetic field
and show in detail how the orbital effect suppresses the critical
current. Finally, when normal current and supercurrent are
simultaneously present in the junction, the distribution function
develops a spatially inhomogeneous component that can be interpreted
as an effective temperature gradient across the junction, with a
sign that is controllable by the supercurrent.  Taken as a whole,
these experiments illustrate the richness and complexity of S/N/S
Josephson junctions in nonequilibrium situations.
\end{abstract}

\pacs{74.50.+r, 73.23.-b, 85.25.Am, 85.25.Cp} \maketitle

\section{Introduction}

When a superconducting metal (S) and a normal metal (N) are placed
in contact with each other, the properties of both metals are
modified near the S/N interface.  This effect, called the
superconducting proximity effect, was widely studied in the
1960's.\cite{deGennes} Our microscopic understanding of the
proximity effect underwent dramatic progress in the 1990's as a
result of new experiments performed on submicron length scales,
coupled with theoretical ideas about phase-coherent transport from
mesoscopic physics.  It is now understood that the conventional
proximity effect in S/N systems and the dc Josephson effect in
S/N/S junctions arise from the combination of three ingredients:
Andreev reflection of electrons into holes (and vice versa) at the
S/N interface, quantum phase coherence of electrons and holes, and
time-reversal symmetry in the normal metal. Our new understanding
of the proximity effect in equilibrium situations and in linear
response transport is demonstrated by a wealth of beautiful
experiments \cite{Pannetier:2000} and is summarized in several
theoretical reviews.\cite{LambertRaimondi,BelzigReview}

In the past several years, research in S/N systems has
increasingly focused on nonequilibrium phenomena. Understanding
nonequilibrium situations is more difficult than understanding
near-equilibrium situations, because the electron energy
distribution function in nonequilibrium may be quite different
from a Fermi-Dirac function. In such situations, the behavior of a
specific sample may depend critically on the rates of
electron-electron or electron-phonon scattering. A pioneering work
in this area was the demonstration by Baselmans \textit{et
al.}\cite{BaselmansNature} that the current-phase relation of a
S/N/S Josephson junction can be reversed, producing a so-called
``$\pi$-junction". This effect is produced by applying a voltage
that suitably modifies the form of the distribution function.

This paper presents results of several experiments performed on
S/N/S Josephson junctions with extra leads connecting the N part of
the devices to large normal reservoirs. Samples are made from
polycrystalline thin films of aluminum (S) and silver (N) deposited
by thermal evaporation. Electrical transport is in the diffusive
limit -- i.e. the electron mean free path is much shorter than all
other relevant length scales in the problem, including the sample
length and the phase coherence length.  In these experiments, the
two superconductors are usually at the same potential, referred to
as ground. Different voltages are applied to the normal reservoirs,
which in most cases cause the distribution function in the
structures to deviate strongly from a Fermi-Dirac distribution.

Several of the experiments have been analyzed quantitatively within
the framework of the Usadel equations,\cite{Usadel:70,BelzigReview}
which are appropriate for S/N samples in the diffusive limit. The
equilibrium component of the Usadel equation is a diffusion equation
describing pair correlations in N and S. The nonequilibrium, or
Keldysh, component consists of two coupled Boltzmann equations for
the spectral charge and energy currents. Incorporating inelastic
scattering into the Keldysh equations involves inserting the
appropriate collision integrals; but this procedure has so far been
followed fully in only a few cases. Moreover, the effect of
inelastic scattering on the equilibrium component of the Usadel
equation or the proximity effect on the collision integrals have
never been included self-consistently to our knowledge. More
commonly, researchers analyzing nonequilibrium phenomena solve
either the Keldysh equation without collision integrals, or the
standard Boltzmann equation with collision integrals but without
superconducting correlations, depending on which aspect of the
problem is more important.  At the end of this paper we compare
these approaches as applied to the last experiment discussed in the
paper.

\begin{figure}[ptbh]
\begin{center}
\includegraphics[width=3.4in]{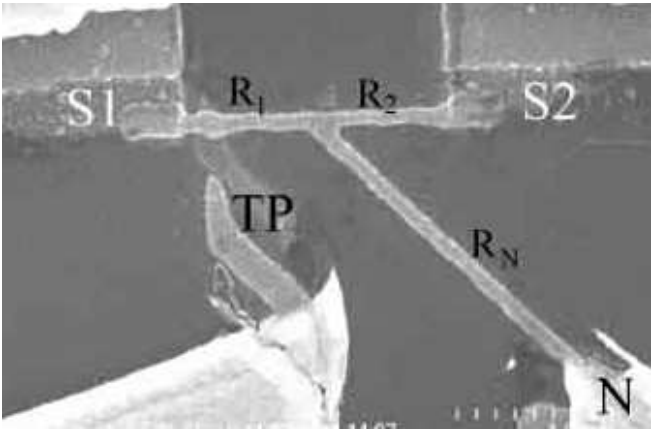}
\end{center}
\caption{SEM image of sample with two superconducting reservoirs,
labeled S1 and S2, and normal reservoir labeled N.  A tunnel
probe, labeled TP, consists of thin Al oxidized prior to
deposition of Ag wire.} \label{Pelt35}
\end{figure}

The paper is organized as follows: Section II describes the sample
fabrication and measurement techniques. Section III describes a
simple experiment, called the ``dangling arm", involving a
3-terminal S/N/S device with a single extra lead to a normal
reservoir.  The dangling arm experiment was first reported by
Shaikhaidarov \textit{et al.}.\cite{Shaikhaidarov:00} We include it
here because it provides a clear demonstration of the superposition
of quasiparticle current and supercurrent in a S/N/S junction, an
essential result for the remainder of the paper. Section IV
describes the $\pi$-junction experiment in 3- and 4-terminal
devices.  The 4-terminal sample allows a direct comparison of the
situations present in the 3-terminal $\pi$-junction\cite{Huang:02}
and the original 4-terminal $\pi$-junction of Baselmans \textit{et
al}.\cite{BaselmansNature} Section V discusses the behavior of the
critical supercurrent as a function of magnetic field applied
parallel to the plane of the sample, and shows the difficulty
involved in trying to achieve a $\pi$-junction by Zeeman splitting
of the conduction electrons.\cite{Tero:00, Yip:2000} The theoretical
calculation relevant to this geometry is given in the appendix.
Section VI discusses an experiment in which supercurrent and
quasiparticle current are independently controlled in a 3-terminal
S/N/S junction, leading to an effective temperature gradient across
the junction.\cite{Tero:03} The local distribution function is
measured by a tunnel probe near one of the S/N interfaces. The
discussion provides information that was not included in our
previous report on this experiment.\cite{Crosser:2006}  Together
these experiments demonstrate the richness of phenomena present in
S/N/S Josephson junctions under nonequilibrium conditions.

\section{Experimental Techniques}
\subsection{Fabrication}
All samples in this work were fabricated using e-beam lithography.
A bilayer of resist was deposited onto an undoped Si wafer covered
only with its native oxide layer. The bilayer was formed by first
depositing a copolymer P(MMA/MAA), followed by a second layer of
PMMA.  The bilayer was exposed by 35-keV electrons and then
developed to make a mask for evaporation. With the resist bilayer,
it is possible to fabricate undercuts in the mask, allowing angled
evaporation techniques to be used.\cite{Dolan:1988} Therefore,
multiple layers of different metals (either $99.99\%$ purity Al or
$99.9999\%$ purity Ag) were sequentially deposited without
breaking vacuum.

These techniques were used to prepare the sample shown in
Fig.~\ref{Pelt35}.  To create the tunnel probe (TP), 30 nm of Al was
deposited while the sample was tilted 45 degrees, creating an actual
thickness of about 21 nm of Al on the surface.  Next, a mixture of
90\%Ar-10\% O$_2$ gas was leaked into the vacuum chamber to a
pressure of 60 Torr. After 4 minutes, the chamber was evacuated
again, in preparation for the following depositions: For the silver
wires; labeled $R_1$, $R_2$, and $R_N$; 30 nm of Ag was deposited
with the plane of the wafer perpendicular to the evaporation source.
For the superconducting reservoirs, S1 and S2, the sample was tilted
45 degrees and rotated 180 degrees in order to deposit 90 nm of Al
(for a 60 nm thickness). Finally, the sample was rotated another 140
degrees in preparation for a final, thick layer of Ag to be
deposited over the normal reservoir, N. The sample in
Fig.~\ref{fourterminal} followed a similar procedure, except
foregoing the first Al deposition and oxidation steps.

\subsection{Experimental setup}

Samples were measured inside the mixing chamber of a top loading
dilution refrigerator.  All electrical leads to the sample passed
through commercial LC $\pi$-filters at the top of the cryostat and
cold RC filters in the cryostat consisting of 2.2 k$\Omega$
resistors in series and 1 nF capacitors coupled to ground.

Current-voltage characteristics (I-V curves) were obtained through
4-probe measurements across the sample. The current was swept
using a triangle wave and several cycles were collected and
averaged together. Measurements of $dI/dV$ were obtained by adding
a slow ($\sim 1$ mHz) triangle wave pattern to the sine output of
a lock-in amplifier. The lock-in amplifier was operated at low
frequencies (less than 100 Hz) to allow for extrapolation of the
system response to zero frequency. Both the in-phase and
out-of-phase components of the signal were recorded and utilized
in the analysis.

\section{Dangling Arm Experiment}
\label{sec:danglingarm}

The dangling arm experiment was first proposed in the Ph.D. thesis
of S. Gueron,\cite{SophieThesis} although a related geometry was
discussed by Volkov two years earlier.\cite{Volkov:1995} The
experiment is performed on a 3-terminal S/N/S Josephson junction
sample similar to the one shown in Fig.~\ref{Pelt35}, in which the
tunnel probe in the lower left was unused.  We label the three
terminals of the sample S$_1$, S$_2$, and N, and the resistances of
the three arms $R_1$, $R_2$, and $R_N$. (We neglect for the moment
the variation of these resistances due to proximity effect.) One
measures the resistance from N to S$_1$, while leaving S$_2$ open
(dangling). Naively, one might expect the measured resistance
between N and S$_1$ to be equal to $R_N + R_1$.  That result would
imply that the current travels directly from N to S$_1$, which in
turn implies that S$_1$ and S$_2$ are at different voltages. Given
that S$_1$ and S$_2$ are coupled by the Josephson effect, the
relative phase $\phi$ between S$_1$ and S$_2$ then accumulates at
the Josephson frequency, $d\phi/dt = 2eV_{12}/\hbar$. If, however,
the injected current $I$ splits into a piece $I_1$ through $R_1$ and
a piece $I_2$ through $R_2$, such that $I_1 R_1 = I_2 R_2$, then
S$_1$ and S$_2$ will be at the same potential. To avoid having a net
current flowing into the dangling arm, the sample must then provide
a supercurrent $I_S$ from S$_2$ to S$_1$ that exactly cancels the
quasiparticle current $I_2$. In this scenario, $R_1$ and $R_2$ are
effectively acting in parallel, and the measured resistance will be
$R_P \equiv R_N + R_1 R_2/(R_1 + R_2)$.
\begin{figure}[ptbh]
\begin{center}
\includegraphics[width=3.4in]{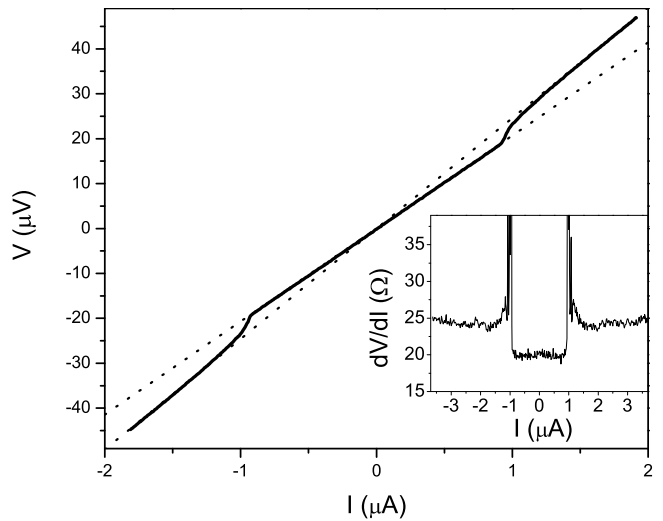}
\end{center}
\caption{Voltage versus current measured between reservoir N and
S$_1$ with S$_2$ floating, at $T=51$ mK.  Dotted lines represent
slopes of 20.7 and 24.6 $\Omega$, which correspond to the
resistances $R_P$ and $R_N + R_1$, respectively.  Inset:
Differential resistance vs. current under similar conditions,
showing agreement between the two measurement techniques.}
\label{DanglingArmRaw}
\end{figure}

Figure \ref{DanglingArmRaw} shows the 2-terminal $I-V$ curve taken
at $T=51$ mK from a sample similar to the one shown in
Fig.~\ref{Pelt35}, with nominal resistance values $R_1 = 7.0$
$\Omega$, $R_2 = 7.0$ $\Omega$, and $R_N = 16.9$ $\Omega$.  The
inset shows $dV/dI$ vs. $I$, providing a clearer view of the
effective resistance. Either plot shows that the resistance is about
20.7 $\Omega$ when the applied current is less than about 0.94
$\mu$A. This resistance is very close to the nominal value of $R_P =
20.4$ $\Omega$. When the current exceeds 0.94 $\mu$A, the resistance
increases to the value 24.6 $\Omega$, which is very close to $R_N +
R_1 = 23.9$ $\Omega$. (Resistance differences less than an Ohm are
attributed to the finite size of the ``T-junction" in the middle of
the sample.) These data confirm the idea outlined in the previous
paragraph, that supercurrent and quasiparticle current can coexist
in the normal region of a S/N/S Josephson junction.

\begin{figure}[ptbh]
\begin{center}
\includegraphics[width=3.4in]{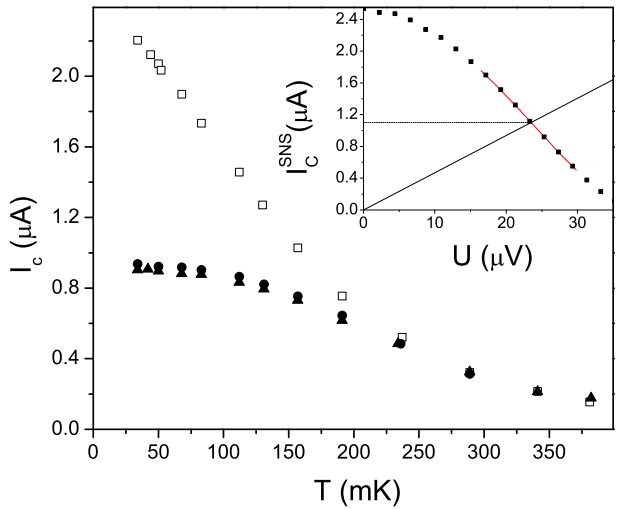}
\end{center}
\caption{Critical current measured between N and S$_1$
($\bullet$), between N and S$_2$ ($\blacktriangle$), and between
S$_1$ and S$_2$ ($\Box$) -- the latter multiplied by the ratio
$(R_1 + R_2)/R_1 \approx 2$ -- for different temperatures. The
three data sets are in close agreement at temperatures above about
250 mK. Inset: Graphical approach to calculation of
low-temperature critical current between N and S$_1$. The dots are
measurements of the critical current across S$_1$-S$_2$, again
multiplied by $(R_1 + R_2)/R_1$, as a function of applied voltage
$U$ between N and S$_1$. The critical current decreases rapidly
with increasing $U$.  The line through the origin represents the
injected current from N. The intersection gives the critical
current $I_c^{NS}$ at the critical value of $U$. Note that all
critical current values in the inset are $15-20 \%$ larger than in
the main panel, due to a small magnetic field B = 125 G present
when the latter data were obtained.} \label{Danglefit}
\end{figure}

The transition at $I_c^{NS}$ where the resistance increases to
$R_N + R_1$ occurs when the supercurrent across the S/N/S
Josephson junction exceeds the S/N/S critical current,
$I_c^{SNS}$. However, since only the fraction $I_2 = IR_1/(R_1 +
R_2) \approx I/2$ of the injected current must be cancelled by the
supercurrent, one should expect that $I_c^{NS} = I_c^{SNS}(R_1 +
R_2)/R_1$. The data in Figure \ref{Danglefit} show that this
expectation is fulfilled at relatively high temperatures, but that
at lower temperature $I_c^{NS}$ falls well below this value.

Two reasons for the small values of $I_c^{NS}$ at low temperature
were given by Shaikhaidarov \textit{et
al.}.\cite{Shaikhaidarov:00} Those authors solved the Usadel
equation analytically in the limit where the S/N interfaces have
high resistance (poor transparency), so that proximity effects are
small and the Usadel equation can be linearized.  They pointed out
that $I_c^{SNS}$ is suppressed below its equilibrium value due to
the applied voltage at N, a result we will reinforce below. They
also argued that the phase-dependence of the resistances $R_1$,
$R_2$, and $R_N$ due to proximity effect causes the measured value
of $I_c^{NS}$ to be smaller than the nominal value $I_c^{NS} =
I_c^{SNS}(R_1 + R_2)/R_1$.

We believe that the effect related to the phase dependence of the
resistances is small and the relative decrease in $I_c^{NS}$ at
low temperature is due predominantly to the decrease in
$I_c^{SNS}$ as a function of $U$. This effect is demonstrated
graphically in the inset to Fig.~\ref{Danglefit}. There the
critical current of the Josephson junction, $I_c^{SNS}$,
multiplied by the constant ratio $(R_1 + R_2)/R_1$, is plotted as
a function of the voltage $U$ applied to the normal reservoir.  As
can be seen in the inset, $I_c^{SNS}$ decreases rapidly as a
function of $U$. The straight line through the origin in the inset
represents the current injected into the sample from the N
reservoir, $U/R_P$, where the resistances are evaluated at phase
difference $\pi/2$ between S$_1$ and S$_2$. The intersection of
the two curves shows the value of the dangling arm critical
current $I_c^{NS}$ (ordinate) at the critical voltage $U_c^{NS}$
(abscissa).  The figure demonstrates the large reduction in S/N/S
critical current due to the applied voltage $U$, which explains
why $I_c^{NS}$ is much smaller than $I_c^{SNS}(R_1 + R_2)/R_1$ at
low temperature. At high temperatures $T\gtrsim{}eU/k_B$, the
relative reduction is less significant due to two reasons: First,
increasing the temperature decreases the critical current
$I_c^{SNS}$, and thereby also $U_c^{NS}$. Moreover, to observe a
sizable reduction in $I_c^{SNS}(U)$, $|eU|$ has to exceed $k_B T$.

\section{S/N/S nonequilibrium $\pi$-Junction}
\label{sec:snspijunction}
\subsection{Three-terminal $\pi$-junction}

Figure \ref{Danglefit} shows, not surprisingly, that the critical
current $I_c$ of an S/N/S Josephson junction decreases when
quasiparticle current is injected into the junction from a normal
reservoir. Indeed, if the only effect of the injected current were
to heat the electrons in the junction, then one would expect the
critical current to continue decreasing monotonically as a
function of the applied voltage $U$.\cite{Morpurgo:1998} That this
is not the case represents a major discovery in nonequilibrium
superconductivity by Baselmans \textit{et
al.}\cite{BaselmansNature} in 2000. Those authors showed that
$I_c$ first decreases as a function of $U$, but then increases
again at higher $U$.  The explanation\cite{Yip:98,Wilhelm:98} for
this counter-intuitive result consists of two pieces.  First, one
can view the supercurrent in the sample as arising from the
continuous spectrum of Andreev bound states in the normal
metal,\cite{Kulik:1970,Heikkila:02} which carry supercurrent in
either direction, depending on their energy. Second, in the
presence of the applied voltage $U$ the electron distribution
function in the junction is not a hot Fermi-Dirac distribution,
but is closer to a two-step distribution - as long as the short
sample length does not allow electron thermalization within the
sample.\cite{Pothier:97} The two-step distribution function
preferentially populates the minority of Andreev bound states that
carry supercurrent in the direction opposite to the majority,
hence it reverses the current-phase relation in the
junction.\cite{Yip:98,Wilhelm:98} Such a Josephson junction is
called a ``$\pi$-junction", because the energy-phase and
current-phase relations are shifted by $\pi$ relative to those of
standard Josephson junctions.

The original $\pi$-junction experiment of Baselmans \textit{et al.}
was performed in a 4-terminal sample, where voltages of opposite
sign were applied to the two normal reservoirs.  Later, Huang
\textit{et al.}\cite{Huang:02} demonstrated that a $\pi$-junction
can also be obtained in a 3-terminal geometry with a single normal
reservoir, a result predicted by van Wees \textit{et
al.}\cite{vanWees:1991} 10 years earlier.

\begin{figure}[ptbh]
\begin{center}
\includegraphics[width=3.4in]{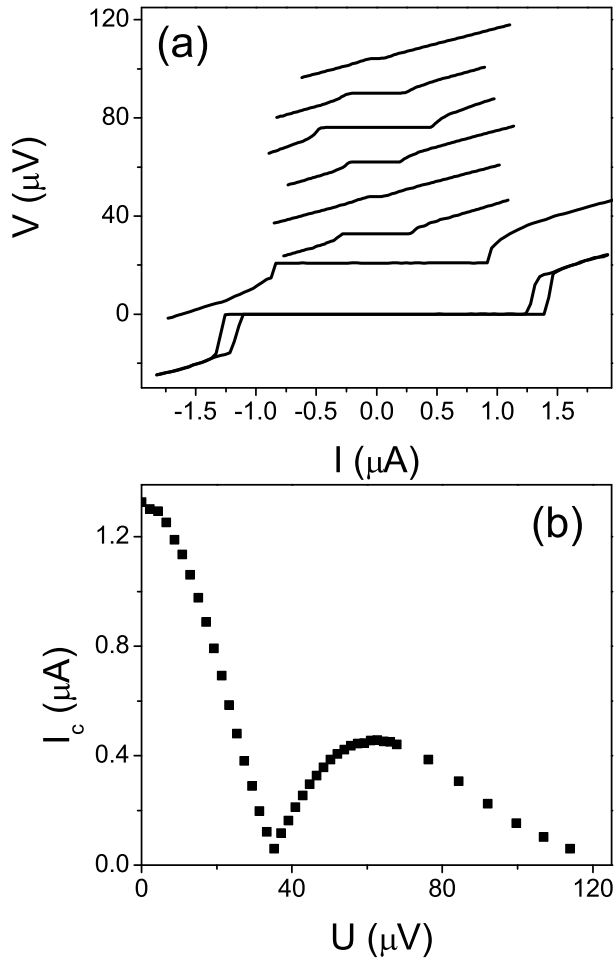}
\end{center}
\caption{a) Voltage vs. current across the S/N/S junction for
selected voltages, $U$, applied to the normal reservoir. Graphs
for different $U$ are offset for clarity, with $U$ = 0, 17, 29,
35, 41, 63, 92, and 114~$\mu$V from bottom to top. The hysteresis
in the $U=0$ data is probably due to heating of the Ag wire in the
normal state. b) Critical current vs. $U$.}
\label{ThreeTermPiJunction}
\end{figure}

Figure \ref{ThreeTermPiJunction}a shows results of a 3-terminal
$\pi$-junction experiment performed on a sample similar to the one
in Fig.~\ref{Pelt35}, where the tunnel probe in the lower left
portion of the figure is not used.  We measure the $I-V$ curve of
the S/N/S Josephson junction using a 4-probe current-bias
measurement, while a dc voltage is simultaneously applied to the
normal reservoir via a battery-powered floating circuit. Figure
\ref{ThreeTermPiJunction}a shows a series of $I-V$ curves at
different values of the voltage $U$ applied to the normal
reservoir. Figure \ref{ThreeTermPiJunction}b shows the critical
current $I_c$ vs. $U$. Notice that $I_c$ initially decreases with
an increasing $U$, as shown in the inset to Fig.~\ref{Danglefit}.
But as $U$ increases further, $I_c$ reaches a minimum value
(indistinguishable from zero in this
experiment\cite{Baselmans:2002}) at $U=U_c \approx 34~\mu$V, then
grows again to reach a second maximum at $U \approx 63~ \mu$V. The
minimum in $I_c$ separates the standard Josephson junction
behavior at low values of $U$ from the $\pi$-junction behavior at
higher $U$. If instead of plotting the critical current $I_c$
(which is by definition a positive quantity) one were to plot the
supercurrent $I_s$ at a fixed phase difference $\phi = \pi /2$
across the junction then the graph would show a smooth curve
passing through zero at $U = U_c$, reaching a local minimum at $U
\approx 63~\mu$V and gradually returning to zero at large $U$.

\begin{figure}[ptbh]
\begin{center}
 \includegraphics[width=3.4in]{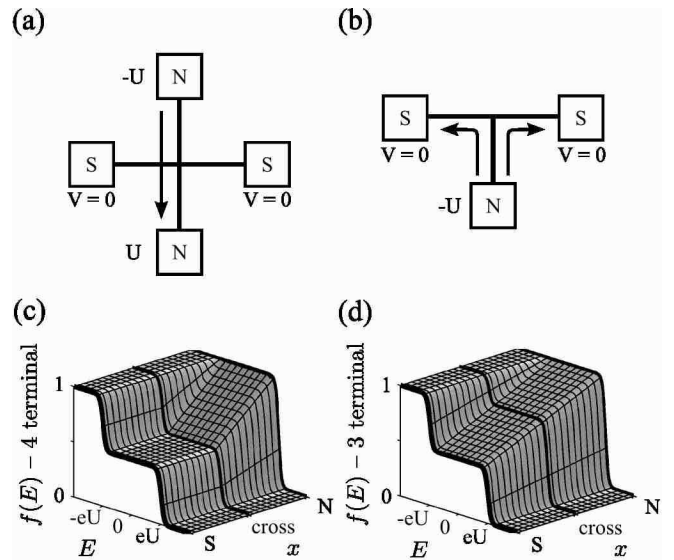}
\end{center}
\caption{a) Depiction of electron flow in four-terminal
configuration.  b) Depiction of electron flow in three-terminal
configuration. c) Schematic representation of the distribution
function on the path between a normal (N) and superconducting (S)
terminal in the structure (a) under high bias $U >> k_B T$. d)
Distribution function in the T-structure (b) under similar
conditions. Due to Andreev reflection $f(\varepsilon)$ is
discontinuous at the N-S interface as explained in the text.}
\label{DistFunctions}
\end{figure}

\subsection{Comparison of four-terminal $\pi$-junctions with symmetric and antisymmetric bias}

The physical explanation of the $\pi$-junction in the 3-terminal
sample is the same as in the 4-terminal sample, with the
differences arising only from the distribution functions. Figure
\ref{DistFunctions} shows a schematic drawing of the distribution
function $f(E)$ along a path from a reservoir N to S for both
4-terminal and 3-terminal samples for $U >> k_B T$, assuming weak
electron-electron interactions in the N wire, and neglecting the
proximity corrections. Notice in figures c and d that $f(E)$
consists of a double-step function, the step height within the
energy range $-eU$ to $eU$ changing with the location along the
wire. As we will show in the next section, the even (in energy)
part of $f(E)$ has no effect on the magnitude of the supercurrent
(in the absence of electron-electron interactions), suggesting
that the voltage-dependent critical current, $I_c(U)$, would be
identical in 3-terminal and 4-terminal samples with identical
dimensions and resistances. However, there are three reasons why
this is not quite true: First, Joule heating is more prevalent in
the 4-terminal device, which rounds the distribution functions
more than in the 3-terminal device. Second, the spectral
supercurrent density, $j_E(E)$, evaluated at the junction point
will be slightly smaller in the 4-terminal sample than in the
3-terminal sample due to the presence of the additional arm
connecting the sample to a normal reservoir.\cite{Heikkila:02}
Finally, $f(E)$ will be slightly more rounded in the 4-terminal
sample due to the increased phase space available for
electron-electron interactions. Roughly speaking, the rate of e-e
interactions at a given energy is proportional to $f(E)[1-f(E)]$,
which is maximized when $f(E) = 1/2$.  Each of these effects serve
to increase $I_c$ in the 3-terminal geometry relative to the
4-terminal geometry.

\begin{figure}[ptbh]
\begin{center}
\includegraphics[width=3.4in]{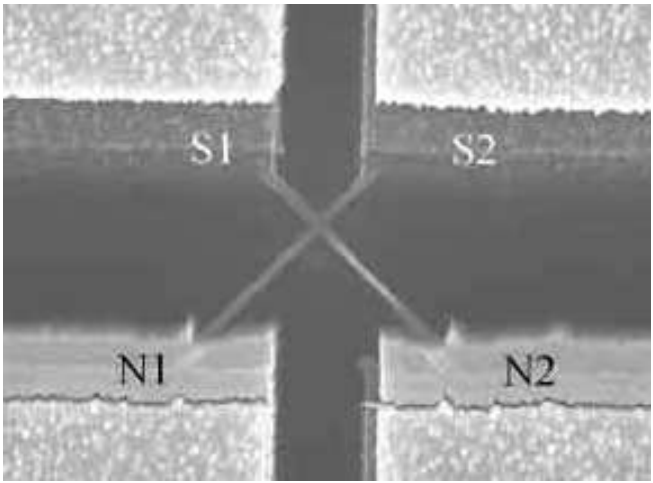}
\end{center}
\caption{SEM image of S/N/S Josephson junction. The `x'-shape is
deposited Ag that connects to (difficult to see) Al reservoirs above
and Ag reservoirs below patterned by angled evaporation.  The
feature in the Ag wire near N2 is likely due to a near burn in the
sample.} \label{fourterminal}
\end{figure}

It is not practical to compare critical currents from two
different samples, since they will never have identical dimensions
nor electrical resistances. Instead, it was proposed in Ref.
\onlinecite{Huang:02} to compare the critical currents in a single
4-terminal sample under conditions of symmetric and antisymmetric
voltage bias of the two normal reservoirs. Figure
\ref{fourterminal} shows the sample we fabricated for this
experiment, with the superconductive reservoirs labeled S1 and S2
and the normal reservoirs labeled N1 and N2.  By applying a
positive potential $U$ to N1 and a negative potential $-U$ to N2,
one reproduces the experiment performed by Baselmans \textit{et
al}. We call this situation antisymmetric bias, since the two
applied voltages differ by a negative sign. In this case the
quasiparticle current overlaps with the supercurrent only at the
crossing point of the sample where the electrostatic potential is
equal to zero.  In contrast, applying the identical voltage $U$ on
both N1 and N2 (with ground defined at one of the superconducting
electrodes), called symmetric bias, will produce a situation
mimicking that in the 3-terminal experiment of Huang \textit{et
al}.\cite{Huang:02}  Notice that by mimicking a 3-terminal sample
with a 4-terminal sample, geometrical differences between the two
experiments are eliminated, so any observed difference in the
critical currents will be due either to e-e interactions or to
Joule heating.

The preceding description of symmetric and antisymmetric biases
holds strictly only if the resistances of the two lower arms are
identical.  Otherwise, application of antisymmetric bias will
result in a nonzero potential at the cross and some quasiparticle
current will flow into the superconducting reservoirs.  In that
case, f(E) will take a form intermediate between those depicted in
Figures \ref{DistFunctions}c and d, which decreases the
measurement contrast between the two biases.  In our experiments
we took care to measure the resistances of all the arms and to
ensure that the voltages at the two normal reservoirs were indeed
equal (for symmetric bias) or opposite (for antisymmetric bias).

\begin{figure}[ptbh]
\begin{center}
\includegraphics[width=3.4in]{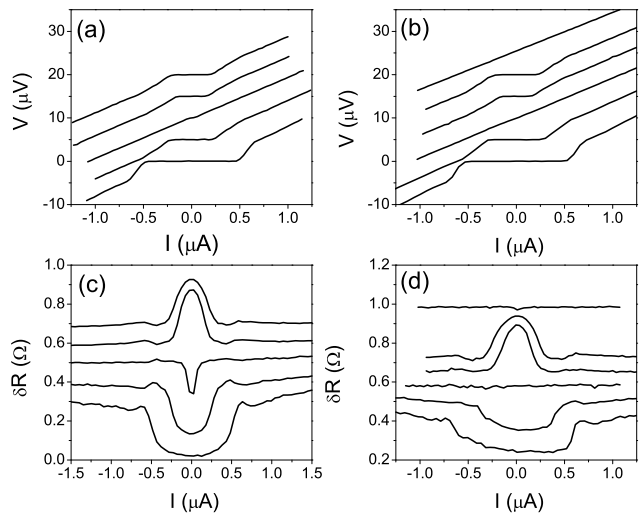}
\end{center}
\caption{Data showing voltage drops across segments of wire either
in antisymmetric or symmetric arrangements while current flows from
S1 to S2. Each line (offset for clarity) corresponds to a different
value of U.  a) Voltage across S1 to S2 for the antisymmetric
measurement.  Applied voltages U are from the bottom: 19, 28, 38,
52, and $71~\mu V$ b) Voltage across S1 to S2 for the symmetric
measurement.  Applied U: 17, 25, 37, 49, 72, and $131~\mu$V. c)
Resistance across N1 to N2 for the antisymmetric measurements. d)
Resistance across N2 to S2 for the symmetric measurements, taken
from voltage measurements in which a constant resistance was
subtracted from the graph.} \label{RawSNS}
\end{figure}

\begin{figure}[ptbh]
\begin{center}
\includegraphics[width=3.4in]{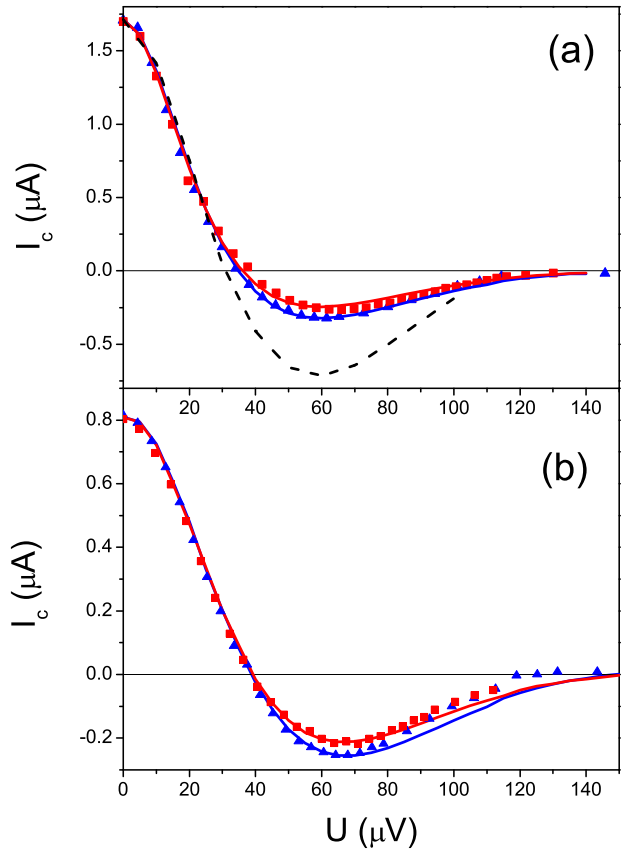}
\end{center}
\caption{(color online) Critical current of a 4-terminal S/N/S
Josephson junction versus voltage $U$ applied to the normal
reservoirs.  The voltages are applied either antisymmetrically
($\blacksquare$) or symmetrically ($\blacktriangle$) to the two
reservoirs.  Solid lines represent simultaneous best fits to data at
different temperatures. Fitting methods are discussed in the text
for data taken at bath temperatures (a) 35 mK and (b) 170 mK.  The
dashed line represents the best fit when Joule heating is excluded.}
\label{fitSNS}
\end{figure}

Figures \ref{RawSNS}a and \ref{RawSNS}b show $I-V$ curves measured
across the S/N/S junction at $T=170$ mK, for several different
values of $U$. Curves with increasing values of $U$ are offset
upward for clarity. Figure \ref{RawSNS}a shows the data for
antisymmetric bias while Fig.~\ref{RawSNS}b shows symmetric bias.
The data follow the same trend observed in
Fig.~\ref{ThreeTermPiJunction}, namely, the critical current first
decreases with increasing $U$, then increases again before finally
disappearing altogether. Figure \ref{fitSNS}b shows this critical
supercurrent as a function of $U$. In this figure we have plotted
the critical current as negative in the range of voltages after
$I_c$ disappears initially, to signify that the junction is in the
$\pi$-state as discussed earlier.

The transition from the 0-state to the $\pi$-state can be
confirmed directly by experiment,\cite{BaselmansNature} without
recourse to the theoretical explanation. The resistances of the Ag
arms of the sample vary with the phase $\phi$ due to the proximity
effect.\cite{Petrashov:95}  The phase $\phi$, in turn, varies
between $\pm \pi/2$ as a function of the supercurrent $I_S$
passing between S$_1$ and S$_2$; hence, one observes a variation
of the resistances as a function of $I_S$. This effect is shown in
Figs. \ref{RawSNS}c and d, in which the resistances between N$_1$
and N$_2$ (N$_1$ and S$_2$) were measured versus $I_S$ for the
antisymmetric and symmetric bias configurations, respectively.
Each curve in the lower two figures has the same value of $U$ as
the corresponding $I-V$ curve in the upper figures.  One can see
that proximity effect induces a local minimum in the resistance at
$I_s = 0$ when the junction is in the 0-state, because that is
where $\phi = 0$. In contrast, the resistance exhibits a local
maximum in the resistance at $I_s = 0$ when the junction is in the
$\pi$-state, because $\phi = \pi$. Interestingly, the top curve in
Fig.~\ref{RawSNS}d shows that at large enough values of $U$, the
system returns to the 0-state since the resistance again shows a
local minimum at $I_s = 0$. This high-$U$ transition from the
$\pi$-state back to the 0-state was not visible in the S/N/S $I-V$
curves.

Figure \ref{fitSNS} shows the behavior $I_c$ vs. $U$ at two
different temperatures.  The squares represent antisymmetric bias
while triangles represent symmetric bias.  Both bias configurations
appear similar in that the samples cross to the $\pi$-state at
nearly the same value of $U$. It should be noted, however, that the
maximum $\pi$-current is larger for symmetric bias than for
antisymmetric bias.  That result is consistent with the qualitative
arguments made above.  In the next section we present a quantitative
analysis of the results.

\subsection{Calculation of the Critical Current in an S/N/S Josephson Junction} \label{ModelSNS}

The amount of supercurrent passing through an S/N/S Josephson
junction may be calculated by \cite{Yip:98}
\begin{subequations}
\begin{eqnarray}
I_S&=&\frac{\sigma_NA}{2}\int^{\infty}_{-\infty}
dE[1-2f(E)]j_E(E)\\
&=&\sigma_NA\int^{\infty}_{0} dE~f_L(E)j_E(E)
\label{Supercurrent}
\end{eqnarray}
\end{subequations}
where $\sigma_N$ and $A$ are the conductivity and cross-sectional
area of the normal metal, respectively. $f(E)$ is the distribution
function within the normal wire, and $f_L(E) \equiv f(-E) - f(E)$
is the antisymmetric component of $f(E)$ with respect to the
potential of the superconductors. The spectral supercurrent
density, $j_E(E)$, is an odd function of energy, and describes the
amount of supercurrent at a given energy travelling between
superconductors with relative phase difference $\phi$. In the
samples considered in the present work, it is generally sufficient
to calculate the supercurrent using the the distribution function
at the crossing point of the wires.

\begin{figure}[ptbh]
\begin{center}
\includegraphics[width=3.4in] {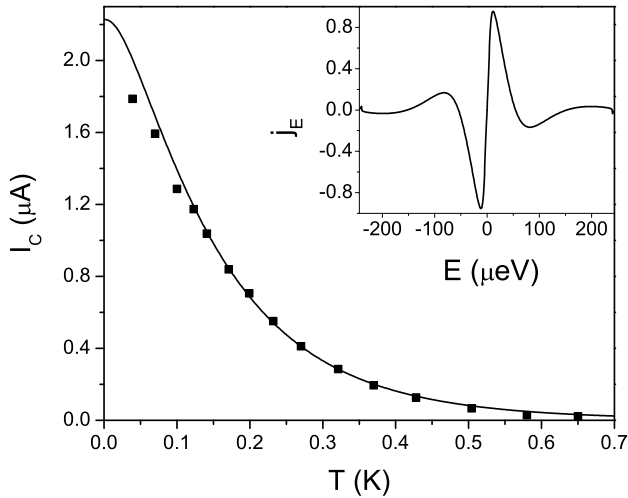}
\end{center}
\caption{Critical current $I_c$ at several temperatures for the
sample shown in Fig.~\ref{fourterminal}. The line is the best fit
by solving equation \ref{Supercurrent} with a Fermi-Dirac
distribution function. Inset: Solution for spectral supercurrent
used to fit data.} \label{fitIcvsT}
\end{figure}

To determine $j_E(E)$, we solve the Usadel equation numerically
using the known physical dimensions and electrical resistances of
the various wire segments of the sample.  We then look for
consistency with the measured temperature dependence of the
equilibrium critical current, $I_c(T)$ shown in
Fig.~\ref{fitIcvsT}. The $j_E(E)$ used to fit these data,
evaluated at $\phi = \pi/2$, is shown in the inset of
Fig.~\ref{fitIcvsT}. Since the length $L$ of the junction is much
longer than the superconducting coherence length $\xi_s$ of the S
electrodes for all the samples studied in this work, the damped
oscillations in $j_E(E)$ occur on an energy scale given by the
Thouless energy, $E_{Th} = \hbar D/L^2$, where $D$ is the
diffusion constant in the wire. $E_{Th}$ characterizes the
temperature scale over which the equilibrium critical current
drops to zero, and also determines the voltage scale $U$ needed to
create a nonequilibrium $\pi$-junction. The transition from the
0-state to the $\pi$-state occurs at $eU \approx 8E_{Th}$.  The
fit shown in Fig.~\ref{fitIcvsT} was obtained with $E_{Th} =
4.11~\mu$eV.

Next we calculate $f(E)$ in the nonequilibrium situation with
antisymmetric bias, i.e. with voltages $U$ and $- U$ applied to
reservoirs N$_1$ and N$_2$. Because we are interested in the
distribution function far from the superconducting reservoirs, we
consider $f(E)$ using the Boltzmann equation.  Let us first ignore
the supercurrent and proximity effects, although inclusion of those
effects will be discussed in detail in Sec.~\ref{TPMsection}. In a
reservoir at voltage $U$, $f(E)$ is a Fermi-Dirac distribution
displaced by energy $eU$, $f(E) =
f_{FD}(E-eU)=[\mathrm{exp}((E-eU)/k_BT)+1]^{-1}$. In the experiment
with antisymmetric bias, we then have $f(E)=f_{FD} (E-eU)$ at
reservoir N$_1$ and $f(E)=f_{FD} (E+eU)$ at reservoir N$_2$. If we
neglect inelastic electron scattering, then in the middle of the
wire (at the intersection of the cross) $f(E)$ has the double-step
shape:
\begin{equation}
f(E) = \frac{1}{2}\left[f_{FD}(E+eU) + f_{FD}(E-eU)\right]
\\\label{doublestep}
\end{equation}
In fact, the odd part of $f(E)$ is the same everywhere in the Ag
wire:
\begin{equation} f_L(E) =
\frac{1}{2}\left[\mathrm{tanh}\left(\frac{E-eU}{2k_BT}\right)+
\mathrm{tanh}\left(\frac{E+eU}{2k_BT}\right)\right] \label{fL}
\end{equation}
This conclusion holds also in the presence of the proximity
effect. At energies $|E| < eU$, the even part, $f_T(E)$, varies
linearly with distance between the two reservoirs (the lower two
arms of the sample), but is zero everywhere along the direct path
connecting the two superconductors (in the ideal case where the
resistances of the two lower arms are equal).

To calculate $f(E)$ in the experiment with symmetric bias, we need
the boundary conditions at the interfaces between the normal wires
and the superconducting reservoirs. For energies below the
superconducting gap, $\Delta$, these conditions are $f_T = 0$ and
${\partial f_L}/{\partial x} = 0$, where $f_T(E) \equiv
1-f(E)-f(-E)$.  These boundary conditions assume high-transparency
interfaces, no charge imbalance in the superconductors, and no heat
transport into the superconductors.\cite{Andreev:64} (Note that
$f_L(E)$ is discontinuous at the N-S interface for energies below
the gap, and returns to the standard form $\tanh(E/2k_B T)$ in the S
electrodes.) The solution for $f(E)$ at the N/S interface is
identical to Eq.~\eqref{doublestep}, but the symmetric component
$f_T(E)$ is nonzero elsewhere in the wire. Notice that the odd
component of the distribution function, $f_L(E)$, is identical in
the two cases everywhere in the sample. The proximity effect induces
a small feature in $f_L(E)$ discussed in Sec.~\ref{TPMsection}, but
it is zero at the crossing point in the middle of the sample.

Calculation of $f(E)$ in the realistic situation requires
consideration of electron-electron interactions in the Ag wire. (The
electron-phonon interaction, in contrast, is much weaker, and need
be considered only in the massive normal reservoirs.  See the
discussion below.)  To incorporate electron-electron interactions,
we solved the Boltzmann equation in the wire numerically, following
previous work by Pierre.\cite{Fredthesis,proximitynote} The results
of this numerical calculation of $f(E)$ in the situations with
either symmetric or antisymmetric bias were extremely similar.
Indeed, the slight additional rounding of $f(E)$ in the
antisymmetric case could not account for the differences observed in
the experiment, shown in Fig.~\ref{fitSNS}.

To account for the difference in the observed $I_c(U)$ between the
two experiments, we next considered the effect of Joule heating on
the temperatures of the normal reservoirs.  (Due to Andreev
reflection at the N/S interfaces, there is no heat transport into
the superconducting reservoirs.) Although we intentionally
fabricated the normal reservoirs much thicker than the wires, this
was not enough to eliminate the effects of Joule heating
altogether. The heat current in a reservoir at a distance $r$ from
the juncture with the wire is given by:
\begin{equation}\label{heatboundarycond}
  \bar{j}^Q = -\pounds \sigma T \nabla T \equiv \frac{P}{\theta rt} \hat{r}
\end{equation}
where $P=I^2R$ is the total power dissipated in the wire, $\sigma$
and $t=310$ nm are the conductivity and the thickness of the
reservoir, respectively, and $\pounds \equiv \pi^2/3 (k_B/e)^2 =
2.44\times 10^{-8}$ V$^2$/K$^2$ is the Lorenz number. (We neglect
the small additional Joule heat generated in the reservoirs
themselves.) The spreading angle $\theta \approx \pi$ if we consider
the combination of the two normal reservoirs shown in
Fig.~\ref{fourterminal}.

Using (\ref{heatboundarycond}) as a boundary condition, one can find
an effective temperature at the wire-reservoir interface equal to
\cite{Nagaev:95}
\begin{equation}\label{Teffective}
  T_{\rm eff}=\sqrt{T^2 + b^2 U^2}
\end{equation}
The temperature far away in the normal reservoir is assumed to be
$T$, the bath temperature.  The factor $b$ is given by
\cite{Henny:99}
\begin{equation}
  b^2=\frac{R_\square}{\theta \pounds R}\mathrm{ln}\frac{r_1}{r_0},
\end{equation}
where $R_\square \equiv 1/(\sigma t)$ is the sheet resistance of
the normal reservoir, $r_0 \approx$ the wire width, and $r_1$ is
the distance over which the electrons in the reservoir thermalize
to the bath temperature via electron-phonon scattering.  The
parameter $b$ varies inversely with the thickness of the metal
reservoir and the electrical resistance encountered by the
quasiparticle current in the wire.  Because the voltage drop $U$
in the antisymmetric bias situation occurs entirely between a
normal reservoir and the crossing point, the resistance $R$ is
smaller than in the symmetric bias situation where $U$ drops fully
from the N reservoirs to the N/S interfaces.  The larger current
in the former case causes more Joule heating, and hence a larger
reservoir temperature.  For our sample, the values of $b$ needed
to fit the data (see solid lines in Figs. \ref{fitSNS}a and
\ref{fitSNS}b) are $2.7$ and $3.2$ K/mV, respectively, for the
symmetric and antisymmetric bias experiments.  Their ratio of 1.2
matches the ratio calculated from the sample parameters.  Their
magnitudes, however, are nearly three times larger than what we
calculate based on the total reservoir thickness.  The experiment
seems to suggest that heat was trapped in the 35-nm Al layer at
the bottom of the reservoirs, rather than immediately spreading
throughout the whole reservoir thickness.\cite{CrosserThesis}

\section{Application of a Parallel Magnetic Field, and the ``Zeeman" $\pi$-junction}
\label{ZeemanSection}

There is a long history of applying magnetic fields perpendicular
to the direction of current flow in
superconductor/insulator/superconductor (S/I/S) Josephson
junctions, to observe the famous Fraunhofer pattern in the
critical current. In S/N/S junctions, the Fraunhofer pattern is
observed only in wide junctions, whereas narrow junctions exhibit
a monotonic decrease of the critical current with field due to the
orbital pair-breaking effect.\cite{angersup}

In this section we discuss the effect of a magnetic field applied
parallel, rather than perpendicular, to the current direction.  In
this geometry there should never be a Fraunhofer pattern. And
because the samples are thin films, one expects the orbital
pair-breaking effect to be much weaker than for a field applied
perpendicular to the plane. In the case of an extremely thin sample
the Zeeman (spin) effect should dominate over the orbital effect of
the field.

The effect of Zeeman splitting on an S/N/S Josephson junction was
studied theoretically in 2000 by Yip\cite{Yip:2000} and by
Heikkil\"a \textit{et al.}.\cite{Tero:00}  Their idea is that the
electronic structure of a normal metal in a large applied magnetic
field resembles that of a weak ferromagnet, in that the up and
down spin bands are displaced by the Zeeman energy. They also
showed how the Zeeman-split junction behaves analogously to the
nonequilibrium S/N/S junction, with the Zeeman energy playing the
role of the voltage in Eqs.~(\ref{Supercurrent}) and (\ref{fL}).
Josephson junctions made with real ferromagnetic materials (S/F/S
junctions) are the subject of intense current interest, as they
can also show $\pi$-junction behavior.\cite{Ryazanov:2001}  Unlike
the $\pi$-junctions discussed earlier in this paper, however, the
$\pi$-junctions in S/F/S systems occur in equilibrium.  They
appear only in particular ranges of the F-layer thickness, due to
spatial oscillations in the superconducting pair correlations
induced in the F metal near the F-S interface by proximity effect.
Those oscillations, in turn, arise from the different Fermi
wavevectors of the spin-up and spin-down electrons in the F metal.
In diffusive S/F/S junctions, the sign of the coupling between the
two superconductors oscillates over a distance scale $\xi_F =
(\hbar D/E_{ex})^{1/2}$, where $D$ is the diffusion constant and
$E_{ex}$ is the exchange energy in the ferromagnet.  In the
standard elemental ferromagnets, $E_{ex}$ is large ($\approx 0.1$
meV), hence $\xi_F$ is extremely short -- on the order of 1 nm.
Control of sample thickness uniformity at this scale is difficult,
hence several groups have used dilute ferromagnetic alloys, with
reduced values of $E_{ex}$, to increase $\xi_F$.  The advantage of
the ``Zeeman" $\pi$-junction is that it is fully tunable by the
field. The disadvantage is that the sample must be thin enough to
minimize the effects of orbital pair-breaking in both the
superconducting electrodes and in the normal part of the junction.

\begin{figure}[ptbh]
\begin{center}
\includegraphics[width=3.4in] {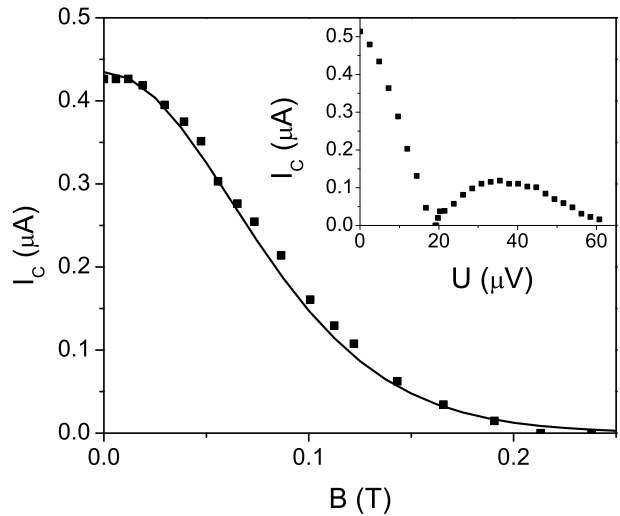}
\end{center}
\caption{Critical current across S/N/S junction as a function of
  external magnetic field applied parallel to the current direction.
  In the absence of orbital pair-breaking effects, the transition into the $\pi$-state
  would be expected near 0.35 T. Markers indicate experimental results and solid
  line the scaling \eqref{eq:magneticfieldscaling} obtained from the
  Usadel equation.  Inset: Critical current versus applied voltage for
  same sample, showing transition to the $\pi$-state at $U=20$ $\mu$V.}
\label{Zeeman}
\end{figure}

Figure \ref{Zeeman} shows a plot of $I_c$ vs. $B$ in an S/N/S
sample whose normal part had length $L = 1.4$ $\mu$m, width
$w~=~50$ nm and thickness $t~=~33$ nm.  The critical current
decreases monotonically to zero, over a field scale of $\approx
0.1$~T.  This result might appear surprising at first glance: At a
field $B = 0.1$ T, the magnetic flux enclosed in the cross-section
of the wire perpendicular to the field is only $\Phi \approx 1.6
\cdot 10^{-16}$ Tm$^2 = 0.08 \Phi_0$, where $\Phi_0 = h/2e$ is the
superconducting flux quantum.  Furthermore, separate tests of the
Al banks confirm that they remain superconducting to fields of
order 0.85 T.

A quantitative understanding of the data in Figure \ref{Zeeman}
can be obtained from a solution to the Usadel equation.  The
analysis discussed in Appendix~\ref{app:usadelmagnetic} shows how
a parallel magnetic field can be absorbed into a spin-flip rate
$\Gamma_{\rm sf}$ in the equations. This allows us to apply the
scaling $I_c(B)/I_c(B=0) \approx \exp(-0.145 \Gamma_{\rm
sf}/E_{Th})$ for the zero-temperature supercurrent of an S/N/S
junction found in Ref.~\onlinecite{hammerup07} and find
\begin{align}\label{eq:magneticfieldscaling}
  I_c(B)/I_c(B=0) &\approx e^{-(B/B_1)^2} \,, \\
  B_1 &\approx 6.43 \frac{\hbar \sqrt{w^2 + t^2}}{eL w t}
      \approx 0.10\;\mathrm{T}
  \,.
\end{align}
Our numerical calculations confirm that this scaling also applies
in our multi-probe experimental geometry. This prediction is in a
good agreement with the experiment, as seen in Fig.~\ref{Zeeman}.

In the limit $w \gg t$, the characteristic field scale $B_1$
varies as $\Phi_0/Lt$, rather than the more intuitive result
$\Phi_0/wt$ we might expect based on the cross-sectional area of
the wire perpendicular to the field.  The physical explanation for
this result was given by Scheer et al.\cite{Scheer:1997} in a
paper discussing universal conductance fluctuations as a function
of parallel field in normal metal wires.  As an electron travels
down the length of a long diffusive wire, its trajectory circles
the cross-section of the wire many times -- on order $N \approx
(L/w)^2$. Because diffusive motion can be either clockwise or
counterclockwise as seen looking down the wire, the standard
deviation in net flux and accumulated phase between different
trajectories is approximately proportional to $B w t \sqrt{N} = B
L t$, which gives the scaling for dephasing.

It is instructive to ask what constraints on the sample geometry
would have to be met to enable observation of the Zeeman
$\pi$-junction.  We estimate the Thouless energy of the sample
discussed in Fig.~\ref{Zeeman} to be $E_{Th} \approx 2.5$ $\mu$eV
both from the temperature dependence of $I_c$ (not shown), and from
the voltage-induced transition to the $\pi$-state at $U_c=20$ $\mu$V
(inset to Fig.~\ref{Zeeman}).  According to theory, the Zeeman
$\pi$-junction should occur when $g\mu_B B \approx 16
E_{Th}$,\cite{Yip:2000,Tero:00}, or $B=0.35$~T.  Attempts to make
thinner samples in order to increase the field scale $B_1$ in
Eq.~\eqref{eq:magneticfieldscaling} were unsuccessful, due to the
tendency of very thin Ag films to ball up.  According to the theory,
much thinner films, with $t/L$ of the order of $0.2
g\mu_B/eD\sim0.001$, will be required to enable observation of the
Zeeman $\pi$-junction.

\section{Engineering the Distribution Function}\label{TPMsection}

The discussion in Section IV.B. implied that the 3-terminal and
4-terminal $\pi$-junctions are similar, with only minor differences
due to a slight decrease in the phase space available for
electron-electron interactions in the 3-terminal case. But that
oversimplified discussion misses some important physics.
Heikkil\"{a} \textit{et al.}\cite{Tero:03} showed that the
superposition of quasiparticle current and supercurrent in the
horizontal wire in the 3-terminal sample induces a change $\delta
f(E)$ in the distribution function at energies of order $E_{Th}$.
The new feature is antisymmetric in space and energy (see
Fig.~\ref{Fig4} for the theoretical prediction and our experimental
results which follow each other nicely), and can be interpreted as a
gradient in the effective electron temperature across the S/N/S
junction. For this reason, the result was dubbed a ``Peltier-like
effect." Although a tiny cooling effect does occur, observing it in
a real electron temperature would require a slightly modified
experimental setup.\cite{Virtanen:07} In the present case, one
should view this effect mostly as a redistribution of the Joule heat
generated in the sample by the applied bias $U$.

In Sec.~\ref{ModelSNS} it was discussed how the distribution
functions behave in the absence of proximity effects and
supercurrent. Including these effects, but ignoring inelastic
scattering, results in the kinetic equations:\cite{Virtanen:04}
\begin{subequations}
\begin{equation}
\frac{\partial j_T}{\partial x} = 0, \; j_T \equiv D_T(x)
\frac{\partial f_T}{\partial x} + j_E f_L +  \emph{T} (x)
\frac{\partial f_L}{\partial x} ;
\end{equation}
\begin{equation}
\frac{\partial j_L}{\partial x} = 0, \;\; j_L \equiv D_L(x)
\frac{\partial f_L}{\partial x} + j_Ef_T - \emph{T}(x)
\frac{\partial f_T}{\partial x} ;
\end{equation}
\label{eq:kinetic}
\end{subequations}
\noindent where $j_T(E)$ and $j_L(E)$ are the spectral charge and
energy currents, respectively.  The energy-dependent coefficients
$D_T$, $D_L$, $j_E$, and $T$ can be calculated from the Usadel
equation \cite{Usadel:70,Virtanen:04}, and vary with the
superconducting phase difference $\phi$ between S$_1$ and S$_2$. In
general, these equations must be solved numerically; however, they
can be solved analytically by ignoring the energy dependence in
$D_T$ and $D_L$ and neglecting the $\emph{T}$ terms. One can show
that, in the presence of both the applied voltage $U$ and a nonzero
supercurrent between S$_1$ and S$_2$, $f_L$ along the horizontal
wire connecting the two superconductors contains a spatially
antisymmetric contribution proportional to $j_E(E)$.  In the exact
numerical solution to Eqs.~\eqref{eq:kinetic}, the feature is
distorted due to the rapid evolution of the diffusion coefficients
$D_T$ and $D_L$ near the N/S interfaces.\cite{Tero:03}

The antisymmetric feature in $f(E)$ can be measured by performing
tunneling spectroscopy with a local superconducting tunnel probe,
which has been demonstrated by Pothier \textit{et
al.}\cite{Pothier:97} to reveal detailed information about $f(E)$ in
a metal under nonequilibrium conditions. In our case, the local
probe must be placed close to the N/S interface where the predicted
feature in $f(E)$ has its maximum amplitude. This location
introduces a new difficulty in our experiment because the density of
states (DOS) of the Ag wire near the N/S interface is strongly
modified by proximity effect. Hence we must first determine the
modified DOS at equilibrium before we measure $f(E)$ under
nonequilibrium conditions.

The current-voltage characteristic of the probe tunnel junction is
\begin{eqnarray} \label{CBconduct}
I(V&)=&-\frac{1}{eR_T}\int dE n_{\rm Al}(E)\int d\varepsilon P(\varepsilon)\\
&\times &[f_{\rm Al}(E)n_{\rm Ag}(E-eV-\varepsilon)(1-f_{\rm Ag}(E-eV-\varepsilon))\nonumber\\
& -& (1-f_{\rm Al}(E)) n_{\rm Ag}(E-eV+\varepsilon)f_{\rm
Ag}(E-eV+\varepsilon)],\nonumber
\end{eqnarray}
where $R_T$ is the normal state tunnel resistance, $n_{\rm Ag}$ and
$n_{\rm Al}$ are the normalized densities of states, and $f_{\rm
Ag}$ and $f_{\rm Al}$ are the electron energy distribution functions
on the Ag and Al sides of the tunnel junction, respectively. The
function $P(\varepsilon)$ characterizes the probability for an
electron to lose energy $\varepsilon$ to the resistive environment
while tunneling across the oxide barrier, an effect known as
``dynamical Coulomb blockade".\cite{IngoldNazarov} $P(\varepsilon)$
was determined from equilibrium measurements at high magnetic field,
where superconductivity is completely suppressed.  Details of the
fitting procedure used to extract $P(\varepsilon)$ were reported
earlier. \cite{Crosser:2006}
\begin{figure}[ptbh]
\begin{center}
\includegraphics[width=3.4in]{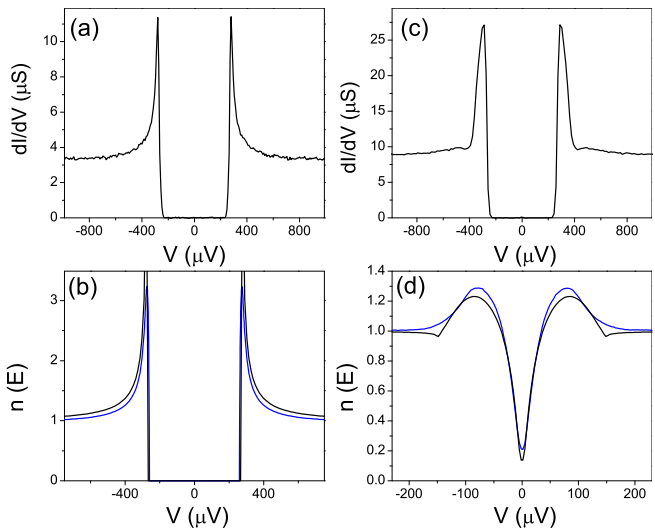}
\end{center} 
\caption{a) Differential conductance data and their best fit for the
reference S/I/N tunnel junction at $B=13$~mT and $T=40$~mK. b) Blue
line is the Al density of states, $n_{Al}(E)$, used to produce the
fit in part (a). (The black line shows the ideal BCS DOS without a
magnetic field, for comparison.) c) The $dI/dV$ data and their best
fit in equilibrium for the tunnel probe on the sample shown in
Fig.~\ref{Pelt35}, at $B=12.5$~mT. d) Blue line is the $n_{Ag}(E)$
used to produce the fit in part (c). The black line is a fit to the
solution of the Usadel equation discussed in the text.}
\label{DOStheory}
\end{figure}

Quantitative analysis of our tunneling data requires an accurate
determination of the superconducting gap, $\Delta$, hence we
fabricated a second S/I/N tunnel junction simultaneously with the
sample, but placed about $20 \mu$m away from it, and with the N side
of the junction far from any superconductor.  Tunneling spectroscopy
measurements on this reference junction, shown in
Fig.~\ref{DOStheory}(a), were fit to Eq.~\eqref{CBconduct} with
$n_{Ag}$ independent of energy and with the standard BCS form for
$n_{Al}$, to provide an accurate determination of $\Delta$.

\begin{figure}[ptbh]
\begin{center}
\includegraphics[width=3.4in]{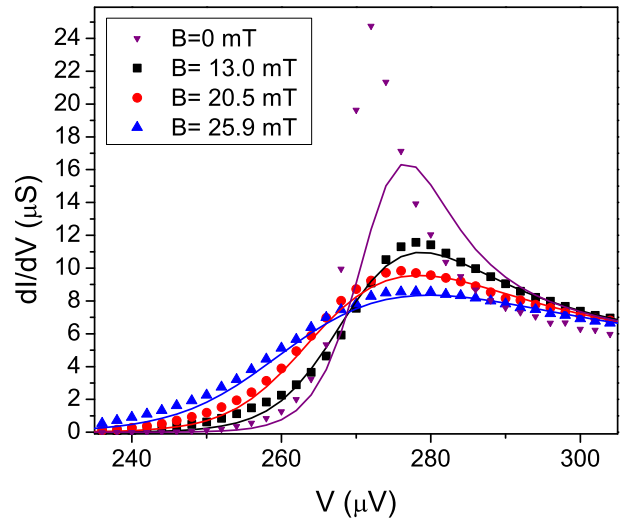}
\end{center}
\caption{(color online) Expanded view of differential conductance
data for $eV$ near $\Delta$ from the tunnel probe far away from
superconducting reservoirs, for select magnetic fields. Symbols
represent data while solid lines are best fits to BCS theory using a
single value of the gap, $\Delta$, and a depairing strength
proportional to $B^2$.  Notice that the data at $B=0$ deviate
significantly from the theory, whereas the data sets with $B>13$~mT
are fit well by the theory.} \label{diffFields}
\end{figure}

Several of our tunnel junctions exhibited sharp anomalies in the
conductance data for voltages close to the superconducting gap;
however, these features disappeared with the application of a small
magnetic field of ~$B=12.5$~mT.\cite{private} Figure
\ref{diffFields} shows $dI/dV$ data for one particular tunnel probe
for different magnetic field strengths. Along with each data set are
fits using the standard BCS form for $n_{Al}(E)$ with a small
depairing parameter proportional to $B^2$, which has been shown to
account well for applied magnetic fields.\cite{anthore:03b} Adding
this term effectively rounds the DOS in the superconductor.
Following the notation of Ref.~\onlinecite{anthore:03b} we
determined a depairing parameter of $\gamma \equiv \Gamma/\Delta
=0.0020$ for $B=12.5$~mT and a superconducting gap in the Al of
$\Delta = 274~\mu$eV. This rather large value for $\Delta$ was
consistent across samples and is believed to be due to oxygen
incorporated into thin, thermally-evaporated Al
films.\cite{dirtyAl,dirtyAl2}

With the form for $n_{Al}(E)$ confirmed, it is possible to analyze
the $dI/dV$ data for the sample tunnel probe, which is in close
proximity to superconducting reservoirs.  Figure \ref{DOStheory}(b)
shows the $dI/dV$ for this probe with an external field of 12.5 mT
applied.  The differences between the $dI/dV$ data from tunnel
probes nearby or far away from superconducting reservoirs arise from
changes in the DOS of the normal wire due to the proximity effect.
Rather than a flat DOS used to fit the data in
Fig.~\ref{DOStheory}(a), the DOS in the Ag wire near the
superconducting reservoirs is modified as shown by the squares in
Fig.~\ref{DOStheory}(d). This shape was obtained by deconvolving the
$dI/dV$ data.

Also shown in Fig.~\ref{DOStheory}(d) is the density of states of
the Ag wire determined from a numerical calculation of the Usadel
equation (solid line). This calculation requires knowledge of the
sample dimensions, the gap in the superconducting reservoirs, and
the Thouless energy.  The sample dimensions were obtained from
scanning electron micrographs, such as the one shown in
Fig.~\ref{Pelt35}.  The gap in the superconducting reservoirs was
found to be $\Delta\approx150~\mu$eV.  This value of $\Delta$ is
much smaller than the value in the Al tunnel probe because the
reservoirs are much thicker than the tunnel probes (and presumably
contain much less oxygen), and because they are close to a normal
metal-superconductor bilayer. Finally, the Thouless energy was
determined by fitting the critical current vs. temperature data as
discussed in Section \ref{ModelSNS}. The value of $E_{Th}$ was then
refined through self-consistent calculations involving both the
finite probe size and the position dependent order parameter
$\Delta$ in the superconducting reservoirs.

\begin{figure}[ptbh]
\begin{center}
\includegraphics[width=3.4in]{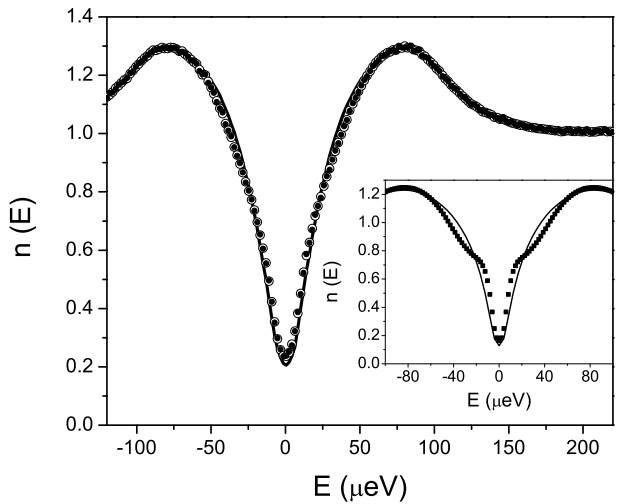}
\end{center}
\caption{a) Density of states of Ag wire at location of tunnel probe
with different amounts of supercurrent flowing across the S/N/S
junction. Solid and hollow circles represent $n_{Ag}$ for
$I_s=0.9I_c$ and $I_s=-0.9I_c$, respectively, while solid line is
for $I_s=0$. Inset: Theoretical results of injecting supercurrent
into device. Solid line for $I_s=0$ and dots for $I_s=\pm .9 I_c$}
\label{DOSequil}
\end{figure}
When supercurrent flows through the S/N/S junction, the phase
difference of the reservoirs, $\phi$, changes $n_{Ag}(E)$. For
this reason, $dI/dV$ data were also taken with supercurrent
flowing across the S/N/S junction. The resulting fits of $n_{Ag}$
for $I=0.9I_c$ and $I=-0.9I_c$, which are identical to each other,
are shown in Fig.~\ref{DOSequil} along with one for $I=0$.  It is
noteworthy that the change of shape (more narrow at low energies,
broader at intermediate ones) is qualitatively consistent with
theoretical calculations shown in the inset.

\begin{figure}[ptbh]
\begin{center}
\includegraphics[width=3.4in]{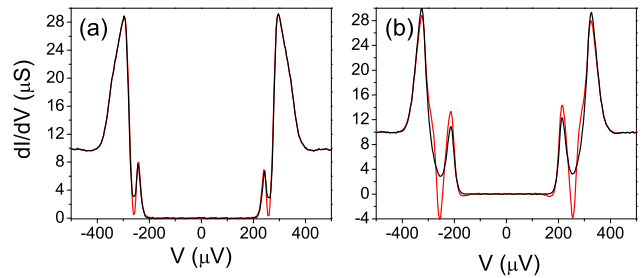}
\end{center}
\caption{Differential conductance tunneling data taken with a
voltage $U$ applied between N to S$_1$ to drive the system out of
equilibrium. Red lines are the best fits using the $n_{Ag}$ data
from Fig.~\ref{DOSequil}. a) $U=22~\mu V$. b) $U=63~\mu V$. The fits
are unable to reproduce the data.} \label{badfit}
\end{figure}
It was anticipated that applying a voltage to the normal lead
would not alter $n_{Ag}(E)$ so that it would be possible to
deconvolve the distribution function, $f_{Ag}(E)$, for the system
out of equilibrium.  Figure \ref{badfit} shows that this
assumption does not hold, as the best fits for applied voltages
$U= 22\mu V$ and $U=63~\mu V$ are poor.  Only by using an altered
$n_{Ag}(E)$ was it possible to fit the data in Fig.~\ref{badfit}.

\begin{figure}[ptbh]
\begin{center}
\includegraphics[width=3.4in]{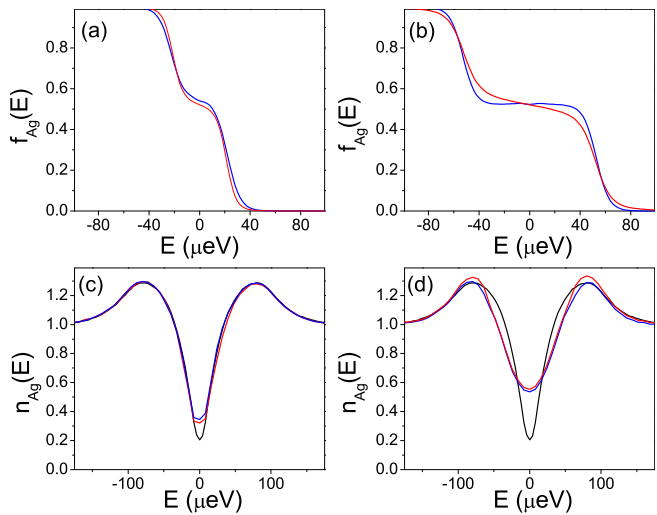}
\end{center}
\caption{Calculated DOS using expected forms for distribution
functions.  a) Distribution functions for $U=25~\mu V$. Blue line
calculated by solving Eqs.~\eqref{eq:kinetic} without collision
integrals. Red line calculated by solving Boltzmann equation
including collisions, but not including superconducting
correlations. b) Distribution functions for $U=63~\mu V$.  Below are
the deconvolved forms for $n_{Ag}$ using the above distribution
functions when c) $U=25~\mu V$ and (d) $U=63~\mu V$. In both, black
lines represent $U=0~\mu V$, for comparison.} \label{DOSstuff}
\end{figure}

\begin{figure}[ptbh]
\begin{center}
\includegraphics[width=3.4in]{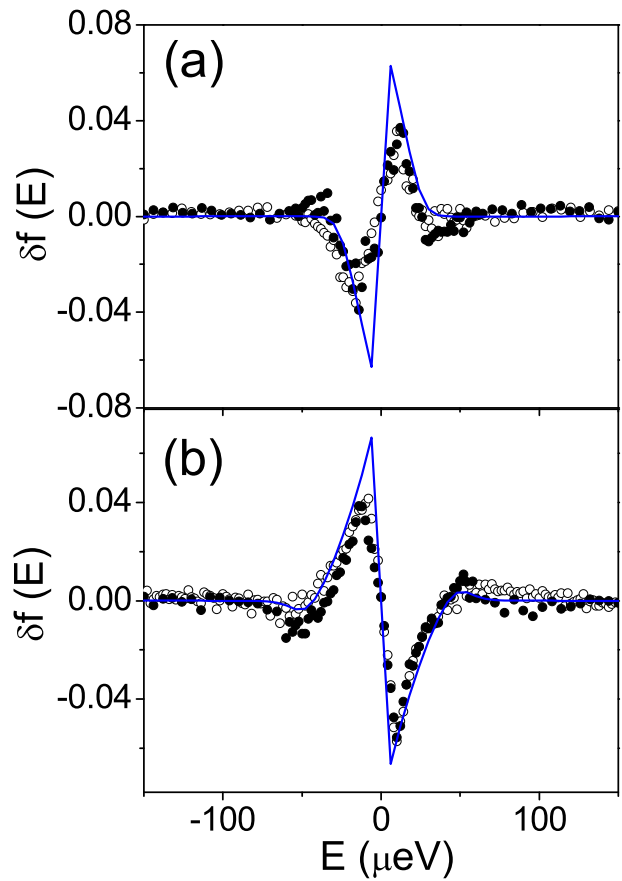}
\end{center}
\caption{(a) $\delta f(E) \equiv f_{Ag}(E,U,I_S) -
f_{Ag}(E,U,-I_S)$ for $U = 22~\mu V$ and $I_S = 0.9 I_c$. (b) Same
quantity for $U = 63~\mu$V and $I_S = 0.9 I_c$, where $U>34~\mu V$
corresponds to the system being in the $\pi$-state. In both
figures, a second data set (open circles) is shown with the signs
of both $U_N$ and $I_S$ reversed. Solid lines are numerical
solutions to Eq.~\eqref{eq:kinetic}.} \label{Fig4}
\end{figure}

Changes in $n_{Ag}(E)$ with increasing $U$ are probably due to a
slight suppression of the gap in the superconducting electrodes,
which are adjacent to superconductor/normal-metal bilayers.  To
estimate how $n_{Ag}(E)$ changes, we used two different forms for
the distribution function $f_{Ag}(E)$. First, we computed
$f_{Ag}(E)$ from Eqs. \eqref{eq:kinetic}(a) and (b), which include
proximity effects due to the nearby superconducting reservoirs, but
neglect inelastic scattering.  Second, we solved the diffusive
Boltzmann equation with collision integrals for electron-electron
scattering, while neglecting superconducting correlations. The two
forms for the distribution functions are shown in Figs.
\ref{DOSstuff}(a) and (b) for two different values of U: $25~\mu V$
and $63~\mu V$. Using those distribution functions, new densities of
states were obtained by deconvolution of the $dI/dV$ data and are
shown as the symbols in Figs. \ref{DOSstuff}(c) and (d). Notice that
the two different forms of the distribution function yield similar
results for $n_{\rm Ag}$. Figures \ref{DOSstuff}(c) and (d) also
show $n_{\rm Ag}$ obtained from equilibrium $dI/dV$ data, as the
solid lines. However, the resulting $n_{\rm Ag}$ does not obey the
sum rule
\begin{equation*}
\int dE (n_{\rm Ag}-1)=0,
\end{equation*}
that should be valid in all situations. We do not know what causes
this discrepancy.

Fortunately it is possible to extract information about
$f_{Ag}(E)$ using a method that is relatively insensitive to the
exact form of $n_{Ag}(E)$, by taking advantage of a near-symmetry
of the data with respect to the direction of $I_S$. The data shown
in Fig. \ref{DOSequil} confirm our expectation that
$n_{Ag}(E,U\!\!=\!\!0,I_S)\!=\!n_{Ag}(E,U\!\!=\!\!0,-I_S)$, a
symmetry that also holds approximately for $U \neq 0$. Hence, one
can analyze the difference between two data sets with opposite
directions of the supercurrent, $dI/dV(V,U,I_S) -
dI/dV(V,U,-I_S)$, which will depend on the differences in the
distribution functions, $\delta f_{Ag}(E) \equiv f_{Ag}(E,U,I_S) -
f_{Ag}(E,U,-I_S)$. The effect of analyzing the data with the wrong
DOS for the Ag is greatly reduced in this case.  The feature we
seek in $\delta f_{Ag}(E)$ is predicted to be odd in $I_s$, hence
it should be the only contribution to $\delta f_{Ag}(E)$. Figure
\ref{Fig4}(a) shows $\delta f$ with $U = 22~\mu$V and $I_S = 0.9
I_c$, which exhibits the predicted feature that is antisymmetric
in energy. The solid lines are the numerical solution to
Eqs.~\eqref{eq:kinetic}, with the parameters $E_{Th}$ and $\Delta$
obtained from the previous fits, with no additional fit
parameters. The computed theory curves agree well with the
experimental data.

A further test of the robustness of the experimental results is to
compare the measured form of $\delta f_{Ag}(E)$ when the signs of
both $U$ and $I_S$ are reversed, i.e. $f_{Ag}(E,-U,-I_S) -
f_{Ag}(E,-U,I_S)$. The results of this second measurement are shown
superimposed on the first in Fig.~\ref{Fig4}(a). The agreement
between the two data sets is excellent.

Interestingly, applying a voltage $U> 34~\mu$V brings this sample
into its $\pi$-state. Figure \ref{Fig4}(b) shows $\delta f(E)$ data
for $U = 63~\mu$V and $I_S = 0.9 I_c$. Compared to
Fig.~\ref{Fig4}(a), the sign of the low-energy feature in $\delta
f(E)$ is reversed, demonstrating that the phase difference $\phi$,
rather than the supercurrent, determines the sign of the new feature
in $f(E)$.

The results of Fig.~\ref{Fig4} indicate that the supercurrent has a
large effect on the electron energy distribution function inside the
normal metal. Such a mechanism has been utilized to explain
\cite{Virtanen:04} the large thermopower measured in Andreev
interferometers \cite{thermopowerexp} --- systems with two
normal-metal and two superconducting contacts. Our results confirm
this mechanism and point out to new phenomena dependent on it, such
as the large Peltier effect \cite{Virtanen:07}: in linear response
to the quasiparticle current (voltage), the supercurrent-induced
change $\delta f$ translates into a change of the electron
temperature and the sign of this change (heating or cooling) depends
on the relative sign of the supercurrent compared to the sign of the
quasiparticle current. One can hence cool part of the structure by
simultaneously applying a quasiparticle current and a supercurrent.

\section{Conclusions}

Superconductor/normal metal hybrid systems exhibit a wealth of
fascinating behaviors, starting with the proximity and Josephson
effects. Driven out of equilibrium, the possibilities increase, from
the non-equilibrium $\pi$-junction to the supercurrent-induced
modification of $f(E)$ discussed in the final section of this paper.
All of these observations can be interpreted with two main concepts:
the spectrum of the supercurrent $j_E(E)$ and the electron
distribution function $f(E)$. The latter can be tuned by applying
voltages or changing the temperature --- the previous for example by
applying a magnetic field. In Secs.~\ref{sec:danglingarm},
\ref{sec:snspijunction} and \ref{TPMsection} we showed in different
schemes how the nonequilibrium $f(E)$ changes the observed
supercurrent and how the supercurrent affects $f(E)$. In
Sec.~\ref{ZeemanSection} we discuss the modifications in $j_E(E)$
due to a magnetic field and the resulting changes in the
supercurrent. To our knowledge, the effect of a parallel magnetic
field on the S/N/S critical current had not been explored in detail
before.

As discussed in Refs.~\onlinecite{Tero:00,Yip:2000}, the Zeeman
effect due to a magnetic field will cause analogous changes in the
supercurrent as the nonequilibrium population of the supercurrent
carrying states. This exact analogy is distorted on one hand due
to the inelastic scattering changing the nonequilibrium
distribution function, and on the other hand the orbital effect
arising from the magnetic field. It remains an experimental
challenge to show this analogy and combine the two effects in the
case when the Zeeman effect dominates over the orbital effect. As
we discuss in Sec.~\ref{ZeemanSection}, the latter would require
constructing extremely thin junctions.

Recently there has been intense interest in the limit where a
Josephson junction behaves as a coherent quantum system with one
degree of freedom.\cite{Bouchiat:98} There is hope that Josephson
junctions will someday provide the building blocks for a quantum
computer.  In the meantime, we hope to have demonstrated that even
in the classical regime, the Josephson junction is full of
surprising new possibilities.

While preparing this manuscript, we learned about recent related
works\cite{angersup} where the magnetic field dependence of the
S/N/S supercurrent was also studied.

Acknowledgements: We thank H. Pothier, D. Esteve, and S. Yip for
many valuable discussions.  This work was supported by NSF grants
DMR-0104178 and 0405238, by the Keck Microfabrication Facility
supported by NSF DMR-9809688, and by the Academy of Finland.

\appendix

\section{Usadel equation and the magnetic field}
\label{app:usadelmagnetic}

The Usadel equation in a magnetic field can be written as follows,
making use of the $\theta$-parameterization $G=\cosh\theta$,
$F=e^{i\chi}\sinh\theta$ of the quasiclassical Green's functions:
\cite{Usadel:70,BelzigReview,Yip:2000,Tero:00}
\begin{gather}
  \label{eq:usadel}
  \hbar D \nabla^2 \theta = -2i(E + \sigma h)\sinh\theta
    + (\hbar\Gamma_{\rm sf} + \frac{v_s^2}{2D})\sinh2\theta
  \,,\\
  \nabla\cdot(v_s\sinh^2\theta) = 0
  \,,
  \quad
  v_s \equiv D[\nabla \chi - 2e\mathbf{A}/\hbar]
  \,.
\end{gather}
Here, $D$ is the diffusion constant, $\mathbf{A}$ the vector
potential, $\Gamma_{\rm sf}$ the spin-flip rate, $v_s$ the
gauge-invariant superfluid velocity and
$h=\frac{1}{2}g\mu\lvert{\mathbf{B}}\rvert$ the Zeeman energy. The
equation is to be solved separately for both spin configurations
$\sigma=\pm$, assuming spin-independent material parameters. The
spin-averaged spectral supercurrent $j_E =
\frac{1}{2}\Im[v_s\sinh^2\theta\rvert_{\sigma=+} +
v_s\sinh^2\theta\rvert_{\sigma=-}]/D$ is obtained from the solutions and
can be used to calculate the observable supercurrent under various
conditions.  Below, we consider these equations in a wire that has an
uniform cross-section $S$, and assume the boundary conditions
\begin{align}
  \chi  &= \pm \phi/2 \,, & \theta &= \theta_0 \,,
  && \text{at $x=0,L$}  \,,
  \\
  \hat{n}\cdot v_s  &= 0 \,, & \hat{n}\cdot\nabla\theta &= 0 \,,
  && \text{on $\partial S$} \,,
\end{align}
where $\partial S$ is the boundary of $S$. These imply that we neglect
details of the current distribution near the terminal--wire contact.

\begin{figure}[t]
  \includegraphics[width=3.4in]{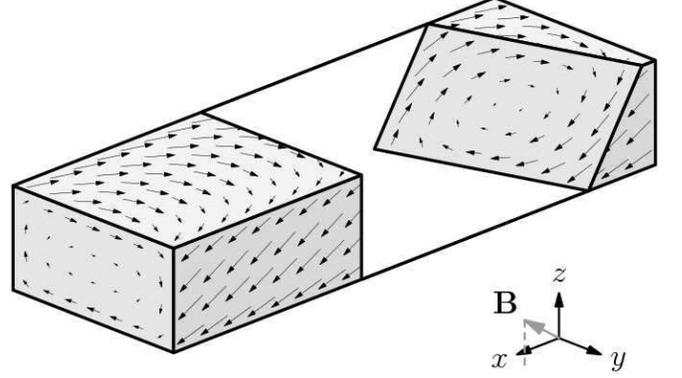}
  \caption{\label{fig:vs} Supercurrent flow induced by a magnetic
    field $\mathbf{B}=(B_x,B_y,B_z)\propto(3,1,2)$ in a thin rectangular
    wire.  The arrows indicate the magnitude and direction of the
    superfluid velocity $v_S$.  Fourth-order variational solution for
    $\chi$ is used here, see text.  }
\end{figure}

When a magnetic field is applied to a wire, in addition to the
Zeeman splitting, the field induces circulating components to the
supercurrent flowing in the wire (see Fig.~\ref{fig:vs}). These
currents contribute to decoherence, for example reducing the
magnitude of the critical current, and in the general case also
prevent reducing Eqs.~\eqref{eq:usadel} to one-dimensional
equations in the direction parallel to the wire.  For thin wires,
however, the additional decoherence can be simply absorbed to the
spin-flip parameter $\Gamma_{\rm{}sf}$ in the one-dimensional
Usadel equation and the vector potential can otherwise be
neglected. \cite{hammerup07,Belzig:96,anthore:03b,deGennesbook}
Below, we show how this conclusion can be reached for an arbitrary
orientation of the magnetic field, and that the results are
consistent with the discussion in Section~\ref{ZeemanSection}.

Reducing Eq.~\eqref{eq:usadel} to a one-dimensional equation is
possible when the transverse dimensions $d$ of the wire satisfy $d
\ll L,l_m$, where $L$ is the distance between the superconducting
contacts and $l_m=\sqrt{\hbar/eB}$ a magnetic length scale. This is
because $\theta$ varies on the length scales of $l_m$ and
$l_E=\sqrt{\hbar D/E} \sim L$ when considering energies $E\sim \hbar
D/L^2$ relevant for the supercurrent. For perpendicular fields it is
also possible to directly choose a proper London gauge where $\chi$
varies slowly in the transverse direction and
$v_S\propto\mathbf{A}$.

Since $l_m\sim{}80\,{\rm nm}$ for $B\sim{}0.1\,{\rm T}$, in the
experimentally interesting situation we have
$w\lesssim{}l_m\ll{}l_E\sim{}L$. To handle the details of the
problem in this case, we apply perturbation theory in the parameter
$\lambda = d/L$.  We choose a coordinate system such that $x$ is the
coordinate parallel to the wire and $y$ and $z$ correspond to
transverse directions, and fix a convenient gauge
$\mathbf{A}=(B_yz-B_zy, -B_x z, 0)$ in which the vector potential is
independent of $x$.  Finally, we rewrite Eqs.~\eqref{eq:usadel} in
the dimensionless variables $\tilde{x}=x/L$, $\tilde{y}=y/\lambda
L$, $\tilde{z}=z/\lambda L$, $\tilde{\mathbf{B}}=eL^2
\mathbf{B}/\hbar\lambda$ and substitute in the regular series
expansion $\theta=\theta_0 + \lambda\theta_1 +
\lambda^2\theta_2+\ldots$, $\chi=\chi_0 + \lambda\chi_1 +
\lambda^2\chi_2+\ldots$.

Requiring the equations corresponding to orders $\lambda^{-2}$,
$\lambda^{-1}$ and $\lambda^0$ of expansion to be separately
satisfied, we first find that the variables $\theta_0$, $\theta_1$ and
$\chi_0$ are independent of $y$ and $z$. We also find that
the first-order response $\delta\chi = \lambda\chi_1$ is given by
\begin{gather}
  \label{eq:chi1eq}
  \nabla_\perp^2 \delta\chi = 0
  \,,\quad
  \hat{n}\cdot \nabla_\perp \delta\chi \rvert_{\partial S} = \hat{n}\cdot 2e\mathbf{A}/\hbar
  \,.
\end{gather}
Here, $\partial S$ is the boundary of the cross-section of the wire,
$\hat{n}$ its outward normal vector, and the operator $\nabla_\perp$
consists of the transverse components of the gradient. This
$x$-independent result applies in the central parts of the wire, away
from boundary layers near the ends of the wire.  Finally, after
averaging the equations of order $\lambda^0$ across the cross section
$S$ of the wire, we arrive at the result
\begin{gather}
  \label{eq:usadel1d}
  \partial_x^2\theta_0 = -2i(E+\sigma h)\sinh\theta_0
       + (\hbar\Gamma_{\rm sf}+\gamma)\sinh2\theta_0
  \,,\\
  \partial_x(\sinh^2\theta_0 \partial_x\chi_0) = 0
  \,,\\
  \label{eq:decoherencegamma}
  \gamma \equiv \frac{\hbar D}{2S}\int_{S}dy\,dz\,(\hat{x}\partial_x\chi_0 + \nabla_\perp\delta\chi - 2e\mathbf{A}/\hbar)^2
  \,,
\end{gather}
This shows how the effective spin-flip parameter is modified by the
applied field $\mathbf{B}$.

The additional decoherence~\eqref{eq:decoherencegamma} depends on the
direction of the field and the cross section of the wire.
For a wire with a circular cross section of radius $R$,
we note that Eq.~\eqref{eq:chi1eq} has the exact solution
$\delta\chi = - eB_x yz/\hbar$. This results to
\begin{gather}
  \gamma =
  \frac{1}{2}\hbar D (\partial_x\chi_0)^2
  + \frac{e^2D}{2\hbar}
  \left(
    \frac{1}{2}R^2B_x^2 + R^2B_y^2 + R^2B_z^2
  \right)
  \,.
\end{gather}
For wires with a rectangular cross section, we cannot solve
Eq.~\eqref{eq:chi1eq} analytically. However, a variational solution is
still possible: we can expand $\delta\chi$ in polynomials of $y$, $z$
to orders $n\le3$ and project Eq.~\eqref{eq:chi1eq} onto this
function basis. From this procedure, we find
\begin{gather}
  \delta\chi \approx -\frac{2eB_x}{\hbar}\frac{d_z^2}{d_y^2 + d_z^2}  y z\,,
  \\
  \gamma =
  \frac{1}{2}\hbar D (\partial_x\chi_0)^2
  + \frac{e^2D}{6\hbar}
  \left(
    \tilde{w}_{yz}^2 B_x^2
    + d_z^2 B_y^2 + d_y^2 B_z^2
  \right)
  \,,
  \\
  \tilde{w}_{yz}^2
  \approx
  \frac{d_y^2 d_z^2}{d_y^2 + d_z^2}
\end{gather}
where $d_y$ and $d_z$ are the width and thickness of the wire. For
orders $n\le4$ we obtain instead
\begin{align}
  \tilde{w}_{yz}^2
  =
  \frac{d_y^2 d_z^2}{d_y^2 + d_z^2}
  \left\{
    1 - \frac{266 d_y^2 d_z^2}{105d_y^4 + 1500 d_y^2d_z^2 + 105 d_z^4}
  \right\}
\end{align}
Approximations using higher-order basis produce only slight
improvements in accuracy to $\gamma$.

We now note that the contribution of the magnetic field to the
decoherence rate $\gamma$ is of the form $e^2Dd^2 B^2/\hbar$ for all
directions of the field, where $d$ is proportional to some
transverse dimension of the wire.  Comparing this to the energy
scale $E_T=\hbar D/L^2$ of the one-dimensional Usadel
equation~\eqref{eq:usadel1d}, we find a dimensionless parameter $(e
B L d/\hbar)^2\propto(\Phi/\Phi_0)^2$ that determines how much the
magnetic field suppresses coherence. Here, the flux $\Phi$
corresponds to an area $L\times{}d$, which is in agreement with the
discussion in Section~\ref{ZeemanSection}.

Finally, note that above we neglected the screening of the magnetic
field by the induced supercurrents. However, this should not be
important in the experimental case, as the Josephson screening
length $\lambda_J=\sqrt{\hbar d^2/2e\mu_0I_cL}\gtrsim200\,{\rm nm}$
is larger than the width of the junction, and the aluminum terminals
are sufficiently thin as to produce only small screening.

\end{document}